# Antiperovskite superconductor LaPd$_3$P with noncentrosymmetric cubic structure

Akira Iyo,[1*] Izumi Hase,[1] Hiroshi Fujihisa,[1] Yoshito Gotoh,[1] Shigeyuki Ishida,[1] Hiroki Ninomiya,[1] Yoshiyuki Yoshida,[1] Hiroshi Eisaki[1], Hishiro T. Hirose[2], Taichi Terashima,[2] Kenji Kawashima[1,3]

[1]National Institute of Advanced Industrial Science and Technology (AIST), Tsukuba, Ibaraki 305-8568, Japan

[2]National Institute for Materials Science, Tsukuba, Ibaraki 305-0003, Japan

[3]IMRA JAPAN Co., Ltd., 2-36 Hachiken-cho, Kariya, Aichi 448-8650, Japan

ABSTRACT Antiperovskites are a promising candidate structure for the exploration of new materials. We discovered an antiperovskite phosphide, LaPd$_3$P, following our recent synthesis of $A$Pd$_3$P ($A$ = Ca, Sr, Ba). While $A$Pd$_3$P and (Ca,Sr)Pd$_3$P were found to be tetragonal or orthorhombic systems, LaPd$_3$P is a new prototype cubic system ($a$ = 9.0317(1) Å) with a noncentrosymmetric space group ($I$-43$m$). LaPd$_3$P exhibited superconductivity with a transition temperature ($T_c$) of 0.28 K. The upper critical field, Debye temperature, and Sommerfeld constant ($\gamma$) were determined as 0.305(8) kOe, 267(1) K, 6.06(4) mJ mol$^{-1}$ K$^{-2}$ f.u.$^{-1}$, respectively. We performed first-principles electronic band structure calculations for LaPd$_3$P and compared the theoretical and experimental results. The calculated Sommerfeld constant (2.24 mJ mol$^{-1}$ K$^{-2}$ f.u.$^{-1}$) was much smaller than the experimental value of $\gamma$ because the Fermi energy ($E_F$) was located slightly below the density of states (DOS) pseudogap. This difference was explained by the increase in the DOS at $E_F$ due to the approximately 5 at.% La deficiency (hole doping) in the sample. The observed $T_c$ value was much lower than that estimated using the Bardeen–Cooper–Schrieffer equation. To explain the discrepancy, we examined the possibility of an unconventional superconductivity in LaPd$_3$P arising from the lack of space inversion symmetry.

INTRODUCTION

Antiperovskites have attracted much attention in recent years due to their high chemical functionality for potential applications arising from their rich material variety.[1] Crystal structures of antiperovskites contain corner-sharing octahedral networks, wherein small-sized elements of groups 13–18 (typically B, C, N, and O) are enclosed inside the octahedra and large cations or anions occupy the spaces surrounded by the octahedra. Perovskite materials have been intensively studied and commercialized for a long time. However, antiperovskite materials have historically remained underdeveloped compared to their perovskite counterparts. Thus, antiperovskites are promising candidates for the development of advanced materials.

Since the discovery of the antiperovskite superconductor MgNi$_3$C,[2] the search for new antiperovskite materials has been actively conducted in the field of superconductivity, and new antiperovskite and antiperovskite-related superconductors have been discovered to date as listed in Table 1. $Li_2M_3B$ can be categolized as an antiperovskite material, in which two small elements fill the space surrounded by the octahedra. Recently, an antiperovskite superconductor—$Mg_2Rh_3P$—was discovered with the same structure type as that of $Li_2M_3B$.[3,4] While searching for the analogue alkaline earth metal–Pt group metal phosphide ternary systems, we successfully synthesized an antiperovskite phosphide $APd_3P$ ($A$ = Ca, Sr Ba) and discovered a superconducting phase hidden in the solid solution between $CaPd_3P$ and $SrPd_3P$, namely $(Ca,Sr)Pd_3P$.[5,6]

In $(Ca,Sr)Pd_3P$, three types of structures (with $Aba2$, $Pnma$, and $I4_1md$ space groups, respectively) appeared as a result of structural phase transitions, depending on the composition and temperature. This is in contrast to the Pt variants $A'Pt_3P$ that have the same structure ($P4/nmm$) irrespective of $A'$. Therefore, we were interested in the synthesis of $LaPd_3P$ and its structure. In this paper, we report the successful synthesis of a new antiperovskite phosphide $LaPd_3P$ and show that $LaPd_3P$ has a cubic structure that has never been observed in any other material including $APd_3P$ or $A'Pt_3P$. In addition, we found that $LaPd_3P$ exhibited superconductivity at approximately 0.3 K and determined its physical properties. We evaluated the electronic band structure of $LaPd_3P$ by first-principles calculations based on crystal structure refinement, and compared the experimental and theoretical results to obtain further understanding of the superconductivity in $LaPd_3P$.

Table 1 Typical antiperovskite and antiperovskite-related superconductors.

| Compounds | $T_c$ (K) | Structure (space group) | Inversion symmetry | Ref. |
|---|---|---|---|---|
| $MgCNi_3$, $CdCNi_3$ | 8 | Cubic ($Pm$-$3m$) | yes | 2,7 |
| $CePt_3Si$ | 0.75 | Tetragonal ($P4mm$) | no | 8,9 |
| $Li_2M_3B$ ($M$ = Pd, Pt) | 8 | Cubic ($P4_132$) | no | 10,11 |
| $ZnNNi_3$ $CuNNi_3$ | 3 | Cubic ($Pm$-$3m$) | yes | 12,13 |
| $A'Pt_3P$ ($A'$ = Ca, Sr, La) | 8.4 | Tetragonal ($P4/nmm$) | yes | 14 |
| $V_3PnN_x$ ($Pn$ = P, As) | 5.6 | Orthorhombic ($Cmcm$) | yes | 15 |
| $SnSr_{3-\delta}O$ | 5 | Cubic ($Pm$-$3m$) | yes | 16 |
| $(Ca,Sr)Pd_3P$ | 3.5 | Orthorhombic ($Pnma$) | yes | 5,6 |
| $SrPd_3P$ | 0.06 | Tetragonal ($I4_1md$) | no | 5,6 |
| $LaPd_3P$ | 0.3 | Cubic ($I$-$43m$) | no | This study |

EXPERIMENTAL SECTION

Material synthesis

A polycrystalline sample was synthesized through a solid-state reaction between a Pd powder (Kojundo Chemical, 99.9 %) and a LaP compound. To obtain LaP, La chips (99.5 %; Furuouchi Chemical Co., Ltd.) and P (99.999 %; Furuouchi Chemical Co., Ltd.) were reacted together in an evacuated quartz tube at 980 °C for 24 h with intermediate grinding. The sample with a nominal composition of $LaPd_3P$ was grounded using a mortar in an $N_2$-filled glove box. The resulting powder was pressed into a pellet (~0.2 g) and subsequently enclosed in an evacuated quartz tube (inner diameter of 8 mm, length of ~70 mm). This sample pellet was then heated at 920 °C for 12 h, followed by furnace cooling. The obtained LaPd$_3$P remains stable in air.

Measurements

Powder X-ray diffraction (XRD) patterns were obtained at room temperature (RT) (~293 K) using a diffractometer (Rigaku, Ultima IV) with Cu$K_\alpha$ radiation. The crystal structures were refined via the Rietveld analysis using the BIOVIA Materials Studio Reflex software (version 2020 R2).[17] Magnetization ($M$) measurements were performed under several magnetic fields ($H$) using a magnetic-property measurement system (Quantum Design, MPMS). The electrical resistivity ($\rho$) in the temperature range of 2–300 K and the specific heat ($C$) in the temperature range of 2–20 K were measured using a physical property measurement system (Quantum Design, PPMS). Resistivity below 2 K was measured using a dilution refrigerator. The composition of the samples was analyzed using an energy-dispersive X-ray spectrometer (Oxford, SwiftED3000) equipped with an electron microscope (Hitachi High-Technologies, TM3000).

Electronic-band structure calculations

First-principles electronic- band structure calculations were performed using the full-potential linearized augmented plane-wave method and generalized gradient approximation for the exchange-correlation potential.[18] Spin-orbit interactions were included in the calculations using a second variational approach. The crystal structure parameters obtained from the experiments in Table 2 were used for the calculations. We used muffin-tin spheres with radii as $r$(La) = 2.5 a.u., $r$(Pd) = 2.44 a.u., and $r$(P) = 1.90 a.u. The plane-wave cutoff $K_{max}$ was selected such that $R_{mt}*K_{max}$ = 7.0, where $R_{mt}$ = $r$(P) is the smallest muffin-tin sphere. We used 1000 and 2000 k-points for the self-consistent field and density of states (DOS) calculations. The calculations were performed using the computer code WIEN2k.[19] The details of the calculations are provided in our previously reported study.[20]

RESULTS AND DISCUSSION

Crystal structure refinement

Figure 1 presents the Rietveld fitting of the XRD pattern of LaPd$_3$P at RT and the XRD pattern in the 10°–60° range with diffraction indices in the inset. The molar composition ratio, La:Pd:P, of the sample was measured to be 0.95(2):3.00(2):1.01(2), which is almost equivalent to the starting composition; however, La was found to be approximately 5 at.% deficient in the sample.

The XRD pattern was fitted assuming a cubic structure with the $I$-43$m$ noncentrosymmetric space group. The fitting with La occupancies as parameters resulted in a 2-3 mol% La deficiency. However, since the reliability factors were almost the same as the fitting with the La occupancies fixed at 1, the presence of the La deficiency could not be verified by the Rietveld analysis. Table 2 summarizes the structural parameters obtained by fitting with the La occupancy fixed at 1. The weighted-profile reliability factor ($R_{wp}$), expected reliability factor ($R_e$), and goodness-of-fit indicator ($S = R_{wp}/R_e$) were 9.22%, 8.85%, and 1.13, respectively, indicating a satisfactory reliability of the analysis.

Figures 2(a)–(c) illustrate the structures of LaPd$_3$P from several directions, and Figure 2(d) shows the structure of LaPt$_3$P for comparison.[14] The structure of LaPd$_3$P (*I-43m*) differs from that of LaPt$_3$P (*P4/nmm*) as well as from that of *A*Pd$_3$P (*I4$_1$md*) and (Ca,Sr)Pd$_3$P (*Pnma*).[5,6] Furthermore, to the best of our knowledge, the structure of LaPt$_3$P has never been reported in any other material, i.e., it is a new prototype structure. Considering that the ionic radii (XII coordination) of La$^{3+}$ (1.36 Å) and Ca$^{2+}$ (1.34 Å) are almost equivalent, the difference in the crystal structure can be attributed to the different valence states. Interestingly, the Pt variants, *A*'Pt$_3$P, have common tetragonal structures (*P4/nmm*) irrespective of *A*'. In LaPt$_3$P, the Pt$_6$P octahedra were asymmetrically elongated along the *c*-axis because of the large alternating shift in P. In LaPd$_3$P, the rotation and tilting of the Pd$_6$P octahedron, in addition to the alternating shift in P, resulted in a different crystal structure. The atomic radii of Pd and Pt (1.40 Å and 1.35 Å, respectively [21]) are close to each other so that it is difficult to explain the change in the structure solely based on the difference in atomic radii. Although Pd and Pt belong to the same group of 10 elements, they have different outer-shell electron configurations, which may be the origin of the observed structural difference between the Pd and Pt variants.

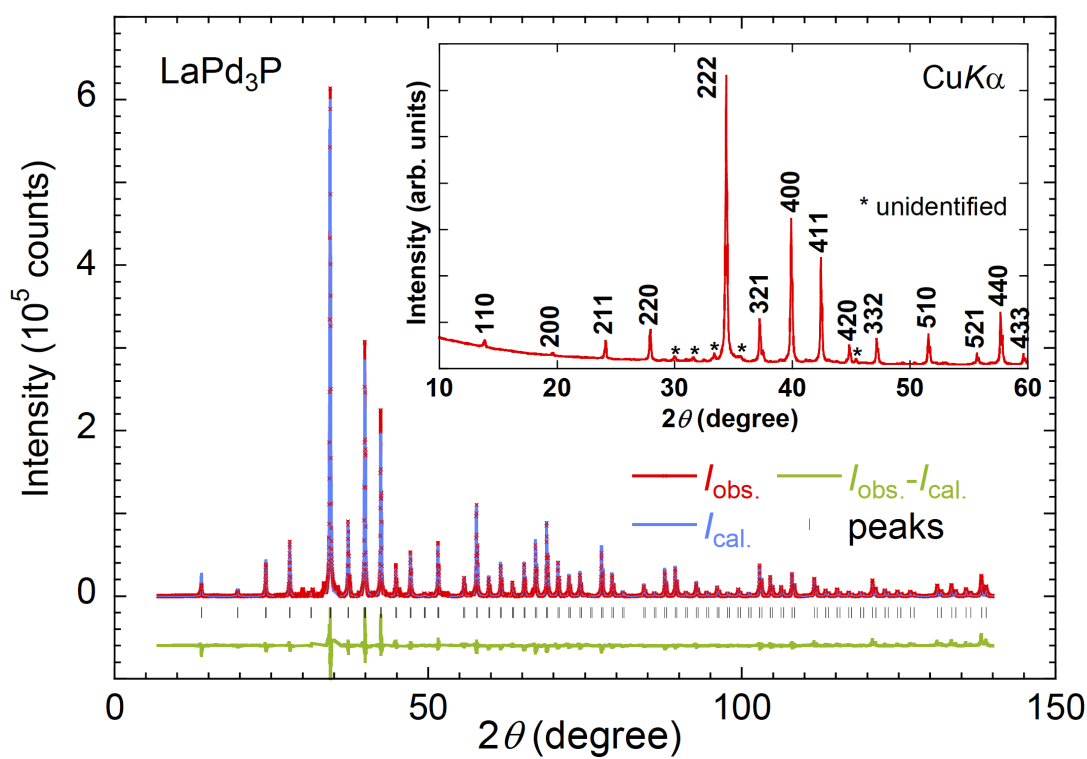

**Figure 1.** Powder XRD pattern and Rietveld fitting for LaPd$_3$P at RT. $I_{obs.}$ and $I_{cal.}$ indicate the observed and calculated diffraction intensities, respectively. The fitting was performed on a background-subtracted diffraction pattern. The inset shows the diffraction pattern for $2\theta = 10°–60°$ with diffraction indices. Unidentified peaks are marked with asterisks.

**Table 2.** Structural parameters obtained by Rietveld refinement for LaPd$_3$P at RT.[a]

| Atoms | Wyckoff positions | $x$ | $y$ | $z$ |
| --- | --- | --- | --- | --- |
| La1 | 2a | 0 | 0 | 0 |
| La2 | 6b | 0.5 | 0 | 0 |
| Pd1 | 24g | 0.2694(1) | $x$ | 0.0472(1) |
| P1 | 8c | 0.3065(3) | $x$ | $x$ |

[a] $I\text{-}43m$ (cubic No. 217); $a$ = 9.0317(1) Å; $V$ = 736.7 Å$^3$ ($Z$ = 8); $R_{wp}$ = 9.22 %; $R_e$ = 8.15%, $S$ = 1.13; The temperature factor was treated as global isotropic and its value was optimized to $U_{iso}$ = 0.025(1). The occupancy for all atoms was fixed at 1.

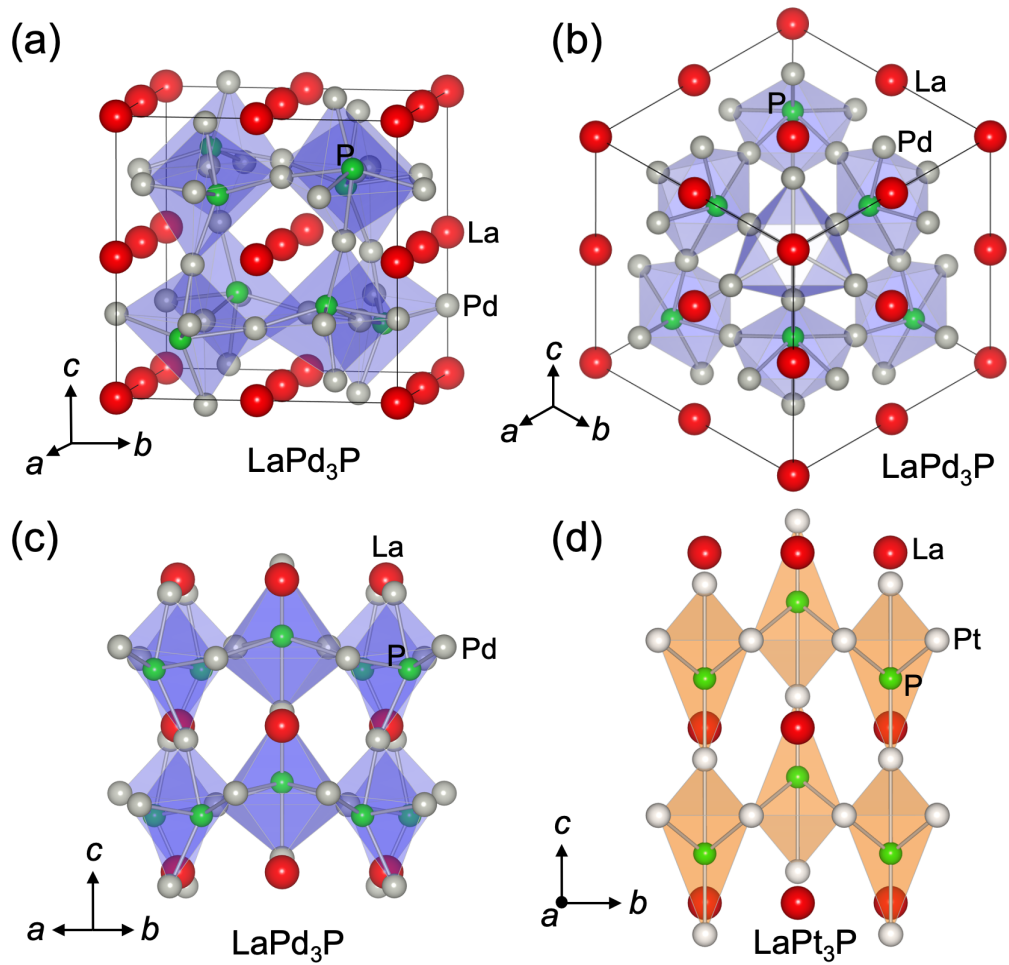

**Figure 2.** (a) Overall view. (b) View from the <111> direction of the crystal structure of LaPd$_3$P. (c) Crystal structure of LaPd$_3$P viewed from the <110> direction. (d) Crystal structure of LaPt$_3$P viewed from the <100> direction.[14] The solid lines indicate unit cells. Crystal structures were drawn using the VESTA software.[22]

Next, we focus on the octahedra in LaPd$_3$P and LaPt$_3$P to discuss their structural differences. Although they are both antiperovskites, the shapes of the Pd$_6$P and Pt$_6$P octahedra are significantly different, as shown in Figures 2(c) and (d). We performed a quantitative structure analysis of each octahedron, and the corresponding results are listed in Table 3. The difference in the octahedral volumes, namely 15.08 Å$^3$ and 15.12 Å3 for LaPd$_3$P and LaPt$_3$P, respectively, was only 1.5 %. However, due to a single unusually long P–Pt bond (2.946 Å) among the six P–Pt bonds of the Pt$_6$P octahedra, the quadratic elongation and bond angle variance [23] were much larger for the Pt$_6$P octahedra than that for Pd$_6$P. The quadratic elongation <$\lambda$> is defined by $<\lambda> = \frac{1}{6}\sum_{i=1}^{6}(\frac{l_i}{l_0})^2$ for the Pd(Pt)$_6$P octahedron, where $l_i$ are the six P-Pd(Pt) bond lengths, and $l_0$ is the center-to-vertex distance of a regular octahedron of the same volume. The bond angle variance $\sigma^2$ is defined by $\sigma^2 = \frac{1}{11}\sum_{i=1}^{12}(\phi_i - \phi_0)^2$ for the Pd(Pt)$_6$P octahedron, where $\phi_i$ are the twelve Pd(Pt)-P-Pd(Pt) bond angles, and $\phi_0$ is the bond angle for a regular octahedron (i.e. 90°). The quadratic elongation <$\lambda$> and the bond angle variance $\sigma^2$ provide quantitative measures of the distortion of the octahedron, independent of the effective size.

Notably, in the absence of distortion in the regular octahedron, the quadratic elongation and bond angle variance were obtained as 1 and 0, respectively. The effective coordination number [24] of P was calculated to be 6 for LaPd$_3$P, while it was approximately 5 for LaPt$_3$P because of the particularly long P–Pt bond. Thus, the structures of LaPd$_3$P and LaPt$_3$P are qualitatively different, and it is appropriate to consider LaPt$_3$P as a compound with a Pt$_5$P pyramidal network. It is interesting to investigate the structural transformations and $T_c$ changes in the solid solution between LaPd$_3$P and LaPt$_3$P.

**Table 3.** Structural analysis of the Pd$_6$P and Pt$_6$P octahedra in the antiperovskite phosphides LaPd$_3$P and LaPt$_3$P, respectively. [a]

| Compounds | LaPd$_3$P | LaPt$_3$P |
| --- | --- | --- |
| Space group | $I$-43m (noncentrosymmetric) | $P4/nmm$ (centrosymmetric) |
| Volume per formula unit (Å$^3$) | 92.08 | 90.70 |
| Octahedral volume (Å$^3$) | 15.06 | 15.12 |
| P–Pd(Pt) bond length (Å) | 2.380×3, 2.389×3 | 2.362×4, 2.519×1, 2.946×1 |
| Quadratic elongation | 1.130 | 1.233 |
| Bond angle variance (deg$^2$) | 337.4 | 750.8 |
| Effective coordination number of P | 6.000 | 5.063 |

[a] Values for structural analysis were obtained using the VESTA software.[22]

Electrical resistivity measurements

Figure 3(a) shows the temperature ($T$) dependence of $\rho$ for LaPd$_3$P. A superconducting transition occurred at the onset and zero resistivity temperatures of 0.35 K and 0.28 K, respectively, as indicated in the inset. The $T$-dependent resistivity exhibited metallic behavior with a strong saturation trend in the normal state. The $RRR$ defined by $\rho(300\ \text{K})/\rho(5\ \text{K})$ was 8.25. The $T$-dependent resistance of LaPd$_3$P differed from those of (Ca,Sr)Pd$_3$P [5,6] and $A$'Pt$_3$P [14]. In particular, it was in contrast to that of LaPt$_3$P, which did not show a clear signature of resistivity saturation up to RT, implying that the electron–phonon coupling and/or electron–electron interactions in LaPd$_3$P were significantly different from those in LaPt$_3$P. To approximate the Debye temperature ($\Theta_D^{\text{res}}$), we fitted $\rho(T)$ above 2 K using the parallel resistance model expressed by: $1/\rho(T) = 1/\rho_{\text{BG}}(T) + 1/\rho_{\text{sat}}$,[25] where $\rho_{\text{BG}}(T)$ and $\rho_{\text{sat}}$ are the Bloch–Grüneisen term and saturation resistivity, respectively. A fit performed using this model, shown by a green curve in Figure 3(a), yielded $\Theta_D^{\text{res}} = 259$ K.

The field dependence of $\rho$ at various temperatures (0.04–1.70 K) is shown in Figure 3(b). The superconducting transitions shifted almost parallel to lower temperature with increasing fields. This behavior was indicative of the bulk superconducting nature of the sample. This was because even if the unidentified phase in the sample indicated by the asterisks in the inset of Fig. 1 was an unknown superconductor, it was highly unlikely that a such a small amount of superconductor exhibited the steep transitions and the parallel shift in magnetic fields. The upper critical field ($H_{c2}$)—defined as the intersection of the line extrapolated from the resistivity and magnetic field axis—is indicated by the dashed line in the main panel. The relationship between $H_{c2}$ and $T$ is plotted in the inset of Figure 3(b); evidently, $H_{c2}$ decreased almost linearly with increasing $T$. Using linear extrapolation, $H_{c2}$ at $T = 0$ K, i.e., $H_{c2}(0)$, was determined as 0.305(8) kOe. The Ginzburg–Landau coherence length $\xi_{\text{GL}}$ was calculated to be $1.04(2) \times 10^3$ Å using $H_{c2}(0) = \Phi_0/2\pi\xi_{\text{GL}}^2$, where $\Phi_0$ is the magnetic flux quantum.

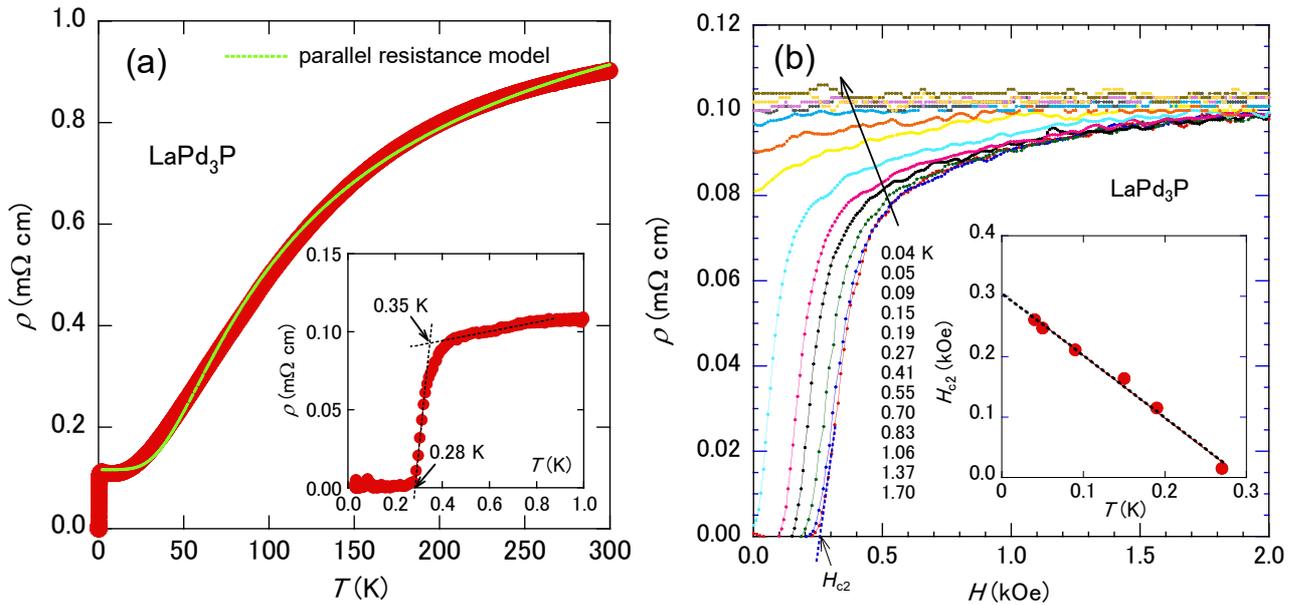

**Figure 3**. (a). Temperature dependence of $\rho$ for LaPd$_3$P. The green curve indicates the fitting using the parallel resistance model. The inset displays a magnified view near $T_c$; (b). $H$ dependence of $\rho$ at temperatures of 0.04–1.70 K. The inset displays temperature dependence of $H_{c2}$. The linear fit of the data is indicated by a dotted line.

Specific heat measurement

Figure 4 shows the $T^2$ dependence of $C/T$ at $H = 0$ for LaPd$_3$P. Notably, the superconductivity-induced jump in the specific heat could not be measured because $T_c$ (~ 0.3 K) was below the lower limit of the measurement range of the apparatus (PPMS). $C/T$ below $T^2 = 50$ K$^2$ was fitted using the expression: $C/T = \gamma + \beta T^2 + \delta T^4$ (red dashed curve in Figure 4), where $\gamma$ is the Sommerfeld constant, and $\beta$ and $\delta$ are related to the phonon contributions to the specific heat. The fitting yielded $\gamma = 6.06(4)$ mJ mol$^{-1}$ K$^{-2}$ f.u.$^{-1}$, $\beta = 0.510(4)$ mJ mol$^{-1}$ K$^{-4}$ f.u.$^{-1}$, and $\delta = 7.8(1) \times 10^{-3}$ mJ mol$^{-1}$ K$^{-6}$ f.u.$^{-1}$ (f.u. stands for formula unit). The Debye temperature ($\Theta_D$) was derived to be 267(1) K using the formula: $\beta = N(12/5)\pi^4 R \Theta_D^{-3}$, where R = 8.314 J mol$^{-1}$ K$^{-1}$ (gas constant) and $N = 5$ (the number of atoms in the formula unit cell). The value of $\Theta_D$ was in good agreement with that obtained from the resistivity analysis ($\Theta_D^{res} = 259$ K). Furthermore, the value of $\Theta_D$ relative to LaPd$_3$P ($T_c = 0.32$ K) was close to that of (Ca$_{0.25}$Sr$_{0.75}$)Pd$_3$P ($T_c = 0.32$ K), i.e., 242 K,[6] while it was higher than that of (Ca$_{0.6}$Sr$_{0.4}$)Pd$_3$P ($T_c = 3.5$ K), namely 184 K [6] and SrPt$_3$P ($T_c = 8.4$ K), namely 190 K.[14] Interestingly, $T_c$ was found to be inversely correlated to $\Theta_D$.

According to the McMillan equation for electron–phonon-mediated superconductors,[26] the electron–phonon coupling constant $\lambda_{e-p}$ is given by: $\lambda_{e-p} = (\mu^* \ln(1.45T_c/\Theta_D) - 1.04)/((1 - 0.62\mu^*)\ln(1.45T_c/\Theta_D) + 1.04)$, where $\mu^*$ is a Coulomb pseudopotential parameter. In this study, we obtained $\lambda_{e-p} = 0.38$ for LaPd$_3$P using $T_c = 0.28$ K, $\Theta_D = 267$ K, and the standard $\mu^*$ value for transition metals (0.13). This relatively small $\lambda_{e-p}$ value indicates that LaPd$_3$P is a weak coupling superconductor.

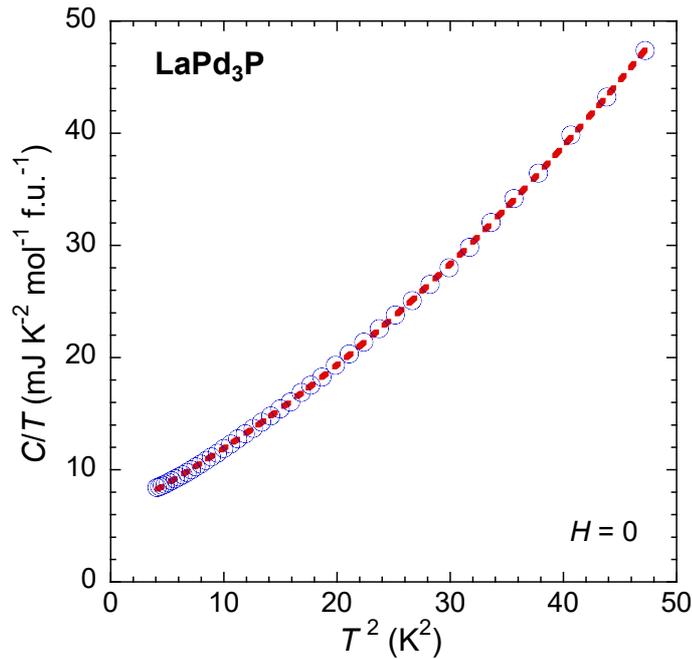

**Figure 4.** $T^2$ dependence of $C/T$ under $H = 0$ for LaPd$_3$P. Dashed curve shows the fitting with $C/T = \gamma + \beta T^2 + \delta T^4$.

Electronic-band structure calculations

The DOS of LaPd$_3$P is shown in Figure 5. We found that the Pd-4$d$ component contributed significantly to most of the DOS of the valence band. A pseudogap just above $E_F$ was the most prominent feature of the DOS of LaPd$_3$P. Consequently, the calculated DOS at $E_F$ ($N(0)^{cal}$) had a small value of 0.950 eV$^{-1}$ f.u.$^{-1}$. This behavior can be explained by a simple ionic crystal model. Assuming the natural ionic states La$^{3+}$ and P$^{3-}$ and imposing a charge neutrality condition, the natural ionic configuration of LaPd$_3$P becomes La$^{3+}$(Pd$^0$)$_3$P$^{3-}$. This means that Pd has a $(4d)^{10}$ configuration; therefore, the Pd-4$d$ band is almost filled and has a band gap. In reality, there exists a moderate hybridization among Pd-4$d$, P-3$p$, and La-5$d$ orbitals, causing the band gap to be the pseudogap.

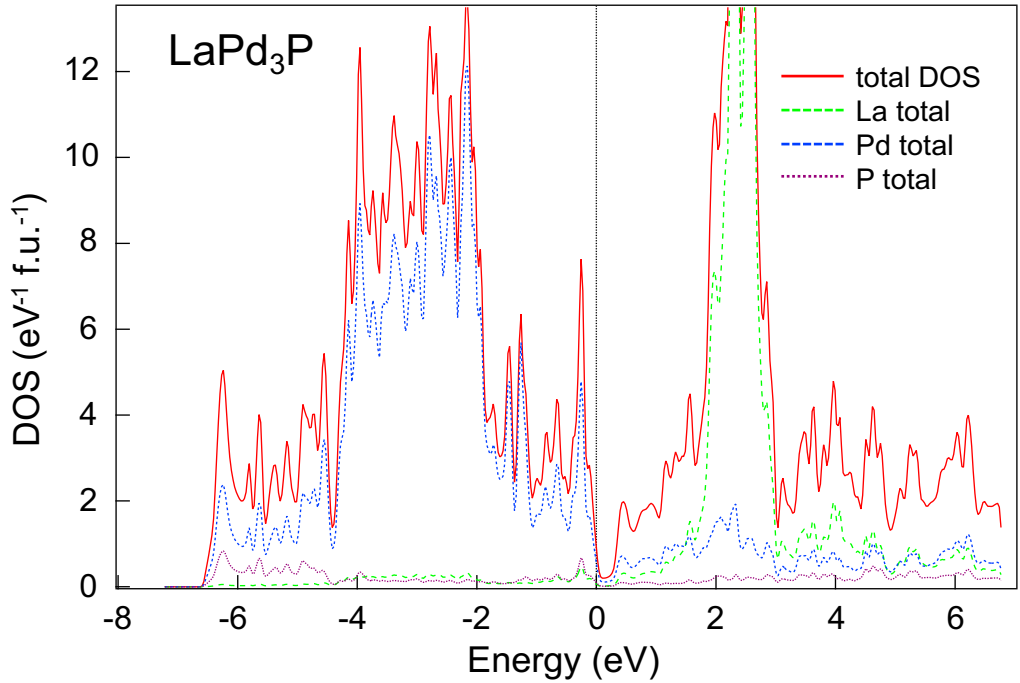

**Figure 5.** DOS of LaPd$_3$P obtained by the first-principles electronic band structure calculations.

The Sommerfeld constant $\gamma^{cal}$ for LaPd$_3$P, calculated from $N(0)^{cal}$, was 2.24 mJ mol$^{-1}$ K$^{-2}$ f.u.$^{-1}$, which was significantly smaller than the experimental value of $\gamma = 6.06$ mJ mol$^{-1}$ K$^{-2}$ f.u.$^{-1}$. The ratio between the experimental and theoretical values of the Sommerfeld constant, $r = \gamma /\gamma^{cal}$, is often referred to as the mass enhancement factor. If there exists a large mass enhancement due to electron–phonon coupling and/or electron–electron interactions, the value of $r$ can even exceed 2. However, for a similar antiperovskite superconductor, namely (Ca,Sr)Pd$_3$P, we found that $\gamma$ and $\gamma^{cal}$ had almost the same value.[20] Therefore, we deduced that the large value of $r$ was not due to this type of mass enhancement; however, the enhancement in $\gamma$ can be attributed to the increase in $N(0)$ due to La deficiency as discussed later.

Figure 5 shows that $E_F$ is at the top of the valence band, and hole doping results in a rapid increase in $N(0)$. The La-5$d$ component of DOS below $E_F$ is very small compared to the Pd-4$d$ component; therefore, as a first approximation, we may consider that La is found in the natural ionic state La$^{3+}$. The composition analysis found a 5 at.% La$^{3+}$ deficiency in LaPd$_3$P. Based on the ionic crystal model described above, the 5 at.% deficient La$^{3+}$ generated 0.15 (= 3×0.05) holes per formula unit, which increases $N(0)^{cal}$ ($\gamma^{cal}$) to 2.69 eV$^{-1}$ f.u.$^{-1}$ (6.34 mJ mol$^{-1}$ K$^{-2}$ f.u.$^{-1}$). This $\gamma^{cal}$ value is consistent with the measured value of $\gamma$ of 6.06 mJ mol$^{-1}$ K$^{-2}$. Thus, the difference between the experimental and theoretical values of $\gamma$ can be explained by the off-stoichiometry of the sample. The parameters obtained from the experiments and the calculations are listed in Table 4.

Next, we theoretically evaluated the value of $T_c$ relative to LaPd$_3$P using a simple Bardeen–Cooper–Schrieffer (BCS) equation $T_c = 1.14\Theta_D \exp(-1/N(0)V)$, where $V$ denotes the effective attractive potential. In our previous study,[20] the range of the $V$ values for the structural analogue, (Ca,Sr)Pd$_3$P, was estimated to be 0.0908–0.115 eV. Hence, we assumed that the $V$ value in LaPd$_3$P falls within this energy range. For LaPd$_3$P, we used $N(0)$ = 2.57 eV$^{-1}$ f.u.$^{-1}$ derived from the measured value of $\gamma$ in lieu of its calculated value. Combining $N(0)$ (2.57 eV$^{-1}$ f.u.$^{-1}$), $V$ (0.0908 - 0.115 eV), and $\Theta_D$ (267 K), we calculated $T_c$ = 4.19–10.3 K for LaPd$_3$P. This value is more than one order of magnitude higher than the observed value of $T_c$ (0.28 K). This result is in contrast to the case of (Ca,Sr)Pd$_3$P, where the experimental $T_c$ values were well reproduced using the same estimation method. If the superconducting mechanism of LaPd$_3$P is equivalent to that of (Ca,Sr)Pd$_3$P, then the $T_c$ of LaPd$_3$P should also be calculated accurately using the same equation. The slight randomness in the sample due to La deficiency should not be the origin of the low $T_c$ because the standard $s$-wave BCS-type superconductors are usually robust against nonmagnetic impurities.

To explain the large difference between the theoretically estimated $T_c$ and the experimental $T_c$, we discuss the possibility that $T_c$ in LaPd$_3$P is suppressed by mechanisms other than the BCS s-wave superconductivity. The crystal structure of LaPd$_3$P lacks space inversion symmetry, implying that the spin-singlet and spin-triplet components can hybridize in the order parameter (superconducting gap). This effect is typically observed in noncentrosymmetric superconductors Li$_2$Pd$_3$B and Li$_2$Pt$_3$B. Li$_2$Pd$_3$B has a much higher $T_c$ value than Li$_2$Pt$_3$B. This is ascribed to the smaller antisymmetric spin-orbit coupling (ASOC) value of Li$_2$Pd$_3$B. This value is determined by both the atomic spin-orbit interaction (SOI) value and the noncentrosymmetric crystal structure.[27,28] Because the crystal structures of Li$_2$Pd$_3$B and Li$_2$Pt$_3$B are almost identical, a smaller SOI value of Pd compared to that of Pt directly results in a smaller ASOC value. Conversely, LaPd$_3$P and (Ca,Sr)Pd$_3$P have different crystal structures; therefore, it is difficult to draw meaningful conclusions from the comparison of the ASOC magnitudes. In spin-triplet superconductivity, the $T_c$ value is easily reduced by nonmagnetic impurities. Therefore, the very low value of $T_c$ obtained in LaPd$_3$P is possibly due to nonmagnetic impurities such as La defects. Recently, the superconducting ground state of LaPt$_3$P was revealed to be a chiral $d$-wave singlet using μSR measurements.[29]

Since the crystal structures of $A$Pd$_3$P ($A$ = Ca, Sr, La) are quite similar to each other, it is plausible to assume that $V$ does not vary much among these materials. Therefore, we believe that $T_c$ is suppressed by ASOC. However, we cannot completely rule out the possibility that $V$ is in fact much smaller than the assumed value. Hence, further experimental and theoretical investigations are required to understand the superconducting state in in LaPd$_3$P.

**Table 4.** Parameters obtained by experiments and calculations for LaPd$_3$P.

| Parameters | |
|---|---|
| $T_c$ (K) ($\rho = 0$) | 0.28 |
| $H_{c2}(0)$ (kOe) | 0.305(8) |
| $\xi_{GL}$ (Å) | 1.04(2) ×10$^3$ |
| $\gamma$ (mJ mol$^{-1}$ K$^{-2}$ f.u.$^{-1}$) | 6.06(4) |
| $\beta$ (mJ mol$^{-1}$ K$^{-4}$ f.u.$^{-1}$) | 0.510(4) |
| $\Theta_D$ (K) | 267(1) |
| $\lambda_{e-p}$ | 0.38 |
| $N(0)^{cal}$ (eV$^{-1}$ f.u.$^{-1}$) | 0.950 [2.69] [a] |
| $\gamma^{cal}$ (mJ mol$^{-1}$ K$^{-2}$ f.u.$^{-1}$) | 2.24 [6.34] [a] |

[a]Values in square brackets were calculated assuming 5 at.% La deficiency.

CONCLUSION

We reported the successful discovery of an antiperovskite superconducting phosphide—LaPd$_3$P—and described its synthesis, crystal structure, physical properties, and electronic band structure. In contrast to the structures of $A$Pd$_3$P and $A'$Pt$_3$P, LaPd$_3$P was found to be a noncentrosymmetric prototype cubic system. In future investigations, it will be necessary to elucidate the mechanism underlying the emergence of such different crystal structures for these antiperovskite phosphides as well as to search for new related antiperovskite (preferably superconducting) materials. The experimental results were discussed based on the comparison with the results of the band structure calculations. The difference between the experimental and theoretical Sommerfeld constants was attributed to the small La deficiency in the sample. We also discussed the possibility of unconventional superconductivity in the noncentrosymmetric LaPd$_3$P as the origin of the discrepancy between the observed and the theoretically estimated $T_c$ values. LaPd$_3$P is also an interesting material to compare with the unconventional superconductor LaPt$_3$P. Further investigations are required to elucidate the superconducting mechanism of LaPd$_3$P.

**Corresponding Author**

iyo-akira@aist.go.jp

ACKNOWLEDGEMENTS

This work was supported by the JSPS KAKENHI (grant numbers JP19K04481, JP19K03731, and JP19H05823) and TIA collaborative research program KAKEHASHI "Tsukuba–Kashiwa–Hongo Superconductivity Kakehashi Project."